\documentstyle[prl,aps,floats,twocolumn,tighten]{revtex}

\def\VEV#1{{\left\langle #1 \right\rangle}}
\def\lsim{\mathrel{\rlap{\lower4pt\hbox{\hskip1pt$\sim$}}
    \raise1pt\hbox{$<$}}}         
\def\gsim{\mathrel{\rlap{\lower4pt\hbox{\hskip1pt$\sim$}}
    \raise1pt\hbox{$>$}}}         

\newcommand{\astpar}[3]{Astropart. Phys.\ {\bf #1}, #3 (#2)}
\newcommand{\plb}[3]{Phys. Lett. B\ {\bf #1}, #3 (#2)}
\newcommand{\physrevd}[3]{Phys. Rev. D\ {\bf #1}, #3 (#2)}
\newcommand{\physrevc}[3]{Phys. Rev. C\ {\bf #1}, #3 (#2)}
\newcommand{\physrevlett}[3]{Phys. Rev. Lett.\ {\bf #1}, #3 (#2)}

\newcommand{\astroph}[1]{astro-ph/#1}

\long\def\comment#1{}

\input{psfig.tex}
\begin{document}
\draft
\twocolumn[\hsize\textwidth\columnwidth\hsize\csname @twocolumnfalse\endcsname
\title{Spin-Dependent WIMPs in DAMA?}
\author{Piero Ullio, Marc Kamionkowski, and Petr Vogel}
\address{Mail Code 130-33, California Institute of Technology,
Pasadena, CA 91125}
\date{October 2000}
\maketitle

\begin{abstract}
We investigate whether the annual modulation observed in the
{\sc Dama} experiment can be due to a weakly-interacting massive
particle (WIMP) with an axial-vector
(spin-dependent; SD) coupling to nuclei.  We evaluate the
SD WIMP-proton cross section under the assumption
that such scattering accounts for the {\sc Dama} modulation, and we do 
the same for a SD WIMP-neutron cross section.  We show 
that SD WIMP-proton scattering is ruled out in a
model-independent fashion by null searches for energetic
neutrinos from WIMP annihilation in the Sun, and that
SD WIMP-neutron scattering is ruled out for WIMP masses $\gtrsim 
20$ GeV by the null result with the {\sc Dama} Xe detector.
A SD WIMP with mass $\lesssim20$ GeV is still
compatible, but only if the SD WIMP-neutron interaction is four
orders of magnitude greater than the WIMP-proton interaction.
\end{abstract}

]

\narrowtext

\section{INTRODUCTION}

Weakly-interacting massive particles (WIMPs) have long been sought
experimentally as the primary
component of the dark halo that enshrouds the Milky Way
\cite{jkg}.  Such dark-matter particles could be detected
directly via observation of the ${\cal O}(10\, {\rm keV})$
nuclear recoils they would produce in a low-background detector
when a dark-matter particle scatters elastically from a nucleus
therein \cite{dirdet,annmod}.  They could also be detected
indirectly via observation of the energetic neutrinos produced
by annihilation of WIMPs that have accumulated in the Sun and/or
Earth \cite{SOS}.

The WIMP can scatter elastically from a nucleus either via a
scalar (spin-independent; SI) interaction, in which case the
WIMP-nucleus cross section scales with the mass of the nucleus
(see, e.g., Fig. 26 in Ref. \cite{jkg}),
or via an axial-vector (spin-dependent; SD) interaction, in which
case the cross section is nonzero only if the nucleus has a
non-vanishing spin.  In this latter case, WIMP-nucleus
scattering occurs primarily via interaction of the WIMP with the 
unpaired proton or neutron, and therefore the WIMP-nucleus
cross section will mostly depend on the
SD WIMP-proton interaction in odd-$Z$ nuclei
and on the SD WIMP-neutron interaction in odd-$N$ nuclei.  

The {\sc Dama} collaboration~\cite{DAMA} has 
reported an annual 
modulation in their 
NaI detector compatible with the
summer-winter variation in the flux of WIMPs incident on the
detector due to the motion of the Earth through the 
halo~\cite{annmod}.  
If interpreted as a WIMP with SI interactions, the effect 
singles out a region in the plane (WIMP mass, WIMP-nucleon cross 
section), centered at about (50 GeV, $7 \times 10^{-6}$ pb), which 
is largely excluded by the null result reported by the {\sc Cdms} 
collaboration using a detector
made of natural germanium~\cite{CDMS}. Further data from more sensitive
detectors will allow a firmer statement on this contradiction. 

An alternative explanation is that the WIMP has a dominantly 
SD coupling with nuclei. In this case, the incompatibility 
between the {\sc Dama} and {\sc Cdms} 
results disappears as Na and I are unpaired-proton nuclei,
while natural germanium contains only $\sim 8\%$ of
the $^{73}$Ge isotope which is an unpaired-neutron nucleus.
Thus, it is conceivable that a WIMP could undergo a SD interaction 
in one detector while remaining invisible to the other.
Such an explanation has been neglected since the required WIMP SD 
cross sections exceed those expected in currently favored 
supersymmetric models.  
However, given the number of free parameters in the minimal
supersymmetric extension of the standard model (MSSM), not to
mention the numerous possible alternatives, a WIMP with
a strong SD coupling to nuclei is certainly plausible.

Here we show
that this possibility can
be rejected experimentally (except for a small unusual corner of
parameter space), rather than through theoretical
prejudice.  WIMPs with SD couplings to protons will 
be captured in the Sun and annihilate therein producing
energetic neutrinos that should be detectable in several existing
neutrino telescopes. Following a model-independent 
approach~\cite{jkgs,katie}, we show that
current upper limits to the 
neutrino flux rule out the possibility that the {\sc Dama} signal 
is due to a WIMP with SD couplings to protons. We check also that the 
alternative hypothesis---that the modulation signal is due to SD 
scattering from neutrons in Na and I---is, 
excluded for WIMP masses $M_\chi \gtrsim20$ GeV by null searches
with odd-neutron targets (the best limit comes from an
experiment with enriched liquid Xe performed by the {\sc Dama}
collaboration as well~\cite{DAMAXe}).  

Indirect-detection rates have already been calculated for a variety of
supersymmetric models that fit the {\sc Dama} modulation signal (see
Ref. \cite{bottino} and references therein). Here, however, we carry 
out a model-independent analysis that does not rely on any details of 
the particle-physics model.

In the next Section, we discuss the procedure to derive the region
in the plane (WIMP mass, WIMP-nucleon cross section) compatible with
the {\sc Dama} modulation signal both for SI and SD interactions. 
Section III reviews the model-independent constraints from 
Ref.~\cite{jkgs} to this parameter space in case of SD interactions 
with protons. Section IV shows that the null result with Xe
strongly constrains the possibility that the {\sc Dama}
modulation could be due to a WIMP SD interactions with
neutrons. Section V concludes.

\section{ANALYSIS OF THE MODULATION SIGNAL}

{\sc Dama} has analyzed the measured modulation assuming it is due to
WIMPs with SI couplings only, publishing the corresponding 
compatible region of WIMP masses and cross sections.
The analysis for WIMPs with SD interactions is completely 
analogous. We now detail the calculation.

The differential direct-detection rate (per unit detector mass)
in a detector made of nucleus $i$ is
\begin{equation}
     \frac{dR_i}{dQ} = \frac{\rho_\chi}{M_{\chi}\,M_{i}}
     \int_{|\vec{v^{\prime}}| \geq v_{\rm{min}}} d^3\vec{v^{\prime}}\;
     f(\vec{v^{\prime}}) \,|\vec{v^{\prime}}|\, 
     \frac{d\sigma_{\chi i}}{dQ} \;,
\label{eq:ddrate}
\end{equation}
where $Q$ is the energy deposited in the detector, and 
$d\sigma_{\chi i}/dQ$ 
is the differential cross section for WIMP elastic scattering with the 
target nucleus.  We assume here that WIMPs of mass $M_{\chi}$  
account for a local dark-matter density $\rho_\chi$ and have a
local (i.e., in the rest frame of the detector) distribution in
velocity space $f$ with normalization $\int
d^3\vec{v^{\prime}} f(\vec{v^{\prime}}) = 1$.

The differential cross section is usually re-written as
\begin{equation}
     \frac{d\sigma_{\chi i}}{dQ} = \frac{1}{Q_{\rm max}} 
     \left(\sigma^{\rm{SI}}_{\chi i} F_i^2(Q) 
     + \sigma^{\rm{SD}}_{\chi i} 
     \frac{S_i(Q,a_p,a_n)}{S_i(0,a_p,a_n)} \right),
\end{equation}
where $Q_{\rm max} = 2 \,m_{\chi i}^2\,|\vec{v^{\prime}}|^2/{M_i}$, 
and $m_{j i}$ is the reduced mass
between particles $j$ and $i$.  
The form-factor suppression of the cross section depends on whether 
the interaction is SI or SD. The SI form factor $F_i(Q)$ is 
relatively simple to calculate; the estimate for the SD form factor
$S_i(Q,a_p,a_n)$ requires more sophisticated nuclear 
modeling~\cite{EngPitVog} and differs in case of SD WIMP-proton
and SD WIMP-neutron scattering.

To compare results obtained with detectors of different materials,
the cross section for WIMP scattering from nucleus $i$ (for SI
or SD) is expressed in terms of the WIMP-proton (neutron) cross
section scaled to zero momentum transfer, 
$\sigma_{\chi i} = (m_{\chi i}^2 / m_{\chi p}^2)
C_{p(n)}\sigma_{p(n)}$.  For SI scattering, $ C_{p(n)}^{\rm SI}
= [Z_i\, f_p + (A_i-Z_i)\, f_n]^2 /f_{p(n)}^2$, where $A_i$ and
$Z_i$ are the atomic mass number and charge of nucleus $i$, and
$f_p\simeq f_n$ are the SI couplings of WIMPs to protons and
neutrons, respectively.  For SD scattering, $C_{p(n)}^{SD} = 4\,
(\lambda_{p(n)}^i)^2 J_i(J_i+1)/3$, where $J_i$ is the total
angular momentum of nucleus $i$, and  $ \lambda_{p(n)}^i =(a_{p} 
\VEV{S_p^i} + a_n \VEV{S_n^i})/(a_{p(n)}J_i)$.  Here, $a_p$ and
$a_n$ are the WIMP SD couplings to protons and neutrons and
$\VEV{S_p^i}$ and $\VEV{S_n^i}$ are the proton and neutron spin 
expectation values in nucleus $i$; they are obtained from nuclear 
models~\cite{EngPitVog}.

Finally, to get the differential detection rate for NaI one has to 
take into account the fact that this is not a monatomic material
and include quenching factors for both Na and I.

Eq.~(\ref{eq:ddrate}) is time dependent; 
the dependence is written implicitly in the 
change of variables between the velocity of the WIMP in the detector 
rest frame and in the galactic frame. The rotation of the Earth
around the Sun gives rise to the well-known annual-modulation 
effect~\cite{annmod}. To a good approximation, the signal event rate 
in the $k$th energy bin can be parameterized by separating a 
component averaged during the year from a time-dependent component,
\begin{equation}
     {\cal{S}}_k = {\cal{S}}_{0,k} + 
     {\cal{S}}_{m,k} \cos\left[2\pi(t-t_0)/T \right],
\end{equation}
where the period $T$ is 1 year and $t_0$ is about June 2.
{\sc Dama} performs a maximum-likelihood analysis and provides values
of ${\cal{S}}_{0,k}$ and ${\cal{S}}_{m,k}$ corresponding to their 
favored region in 4 energy bins, with electron equivalent 
(measured) energies $Q_{\rm ee}$ between 2 and 6 keV~\cite{DAMA}. 

Since we cannot perform an analysis on the raw 
data, we consider an approximate method to decide whether a WIMP
candidate is compatible with the {\sc Dama} modulation or not.  We
follow Ref. \cite{BrhRos} and define the statistical variable,
\begin{equation}
     \kappa = \sum_k \left[
     \frac{\left({\cal{S}}_{0,k}^{\rm th}
     -{\cal{S}}_{0,k}^{\rm exp}\right)^2}
     {\left(\Delta {\cal{S}}_{0,k}^{\rm exp}\right)^2}
     + \frac{\left({\cal{S}}_{m,k}^{\rm th}
     -{\cal{S}}_{m,k}^{\rm exp}\right)^2}
     {\left(\Delta {\cal{S}}_{m,k}^{\rm exp}\right)^2} \right] \,,
\end{equation}
where ${\cal{S}}_{0,k}^{\rm exp}$ is the experimental result and
${\cal{S}}_{0,k}^{\rm th}$ is the theoretical prediction for the 
WIMP candidate being investigated.

Before considering the SD case of interest to us here, we first 
carry out our analysis with a SI WIMP-nucleon coupling to test 
whether our technique reproduces
the results of the full {\sc Dama} analysis that has been published
for this case.  To do so, we evaluate ${\cal{S}}_{0,k}^{\rm th}$ 
and ${\cal{S}}_{m,k}^{\rm th}$ as a function of WIMP mass and
$\sigma_{p}^{\rm{SI}}$ (we assume, as {\sc Dama} did, that $f_p = f_n$;
we take also their choice of form factors and astrophysical
parameters). The $3\sigma$ region singled out by {\sc Dama} with their 
maximum-likelihood method in the ($M_\chi$, $\sigma_{p}^{\rm{SI}}$) 
plane is reproduced fairly well if we require that $\kappa < 60$.

We now turn to our analysis of a WIMP with SD interactions.
We take the form factor for SD scattering from Na and I from 
Ref. \cite{NaIformfactor} and again assume that the WIMP candidate 
is compatible with the {\sc Dama} modulation if $\kappa < 60$.
We consider separately the case of coupling with protons only
($a_{p}\neq 0$, $a_{n}= 0$) and with neutrons only 
($a_{p}= 0$, $a_{n}\neq 0$); the results are indicated by the 
shaded regions, respectively, in Fig. \ref{fig:sdp} and 
in Fig. \ref{fig:sdn}.

\section{A WIMP-PROTON INTERACTION?}

\begin{figure}[t]
\vskip -1.0cm
\centerline{\psfig{file=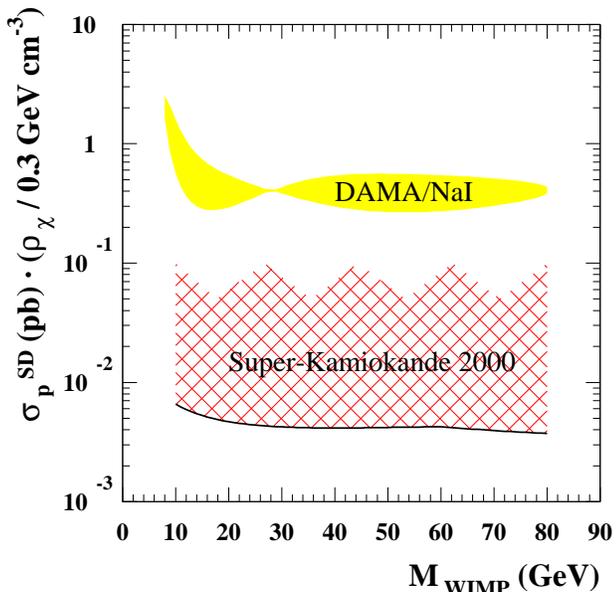,width=3.5in}}
\caption{The shaded region shows the parameter space (in WIMP
         mass versus SD WIMP-proton cross section)
	 implied by the {\sc Dama} annual modulation for a WIMP with
	 exclusively SD interactions with protons
	 and no interaction with neutrons.  The solid curve
	 indicates the upper bound to the SD
	 WIMP-proton cross section from null searches for
	 neutrino-induced upward muons from the Sun;
         thus the cross hatched region is excluded.} 
\vskip -0.2cm
\label{fig:sdp}
\end{figure}

We examine first the case of WIMPs with couplings with protons.
Such WIMPs scatter efficiently from protons in the Sun,
losing enough energy to become gravitationally bound to the
Sun. This process creates an enhancement in the density of
WIMPs at the center of the Sun.
The WIMPs can then annihilate in the
Sun, and among the decay products of the annihilation products
will be energetic neutrinos (energies of order half the WIMP
mass) that can escape the Sun.  Such neutrinos can produce
upward muons via a charged-current interaction in the material
below a neutrino detector, such as that at IMB, Super-Kamiokande,
Baksan, MACRO, and/or AMANDA.  

The calculation of the flux of such neutrino-induced upward
muons from the Sun is straightforward. For the masses we are
considering, the flux can be written
\cite{jkg,jkgs}\footnote{Note in Eq. (9.55) in Ref. \cite{jkg},
the factor $\tanh(t_\odot/\tau_\odot)$ should be squared and
there should be a factor of $\xi(m_\chi)$ on the right-hand side.}
\begin{eqnarray}
     \Gamma  &= &0.016\,{\rm m}^{-2}{\rm yr}^{-1}\, (M_\chi/{\rm
     GeV}) (\sigma^{\rm{SD}}_{\chi p}/ 10^{-40}\, {\rm cm}^{2}) 
     \nonumber \\
     && (\rho_{\chi}/ 0.3 \,{\rm GeV}\,{\rm cm}^{-3}) 
     S(m_\chi/m_H) \xi(m_\chi),
\label{eqn:indirectrate}
\end{eqnarray}
Here, $S(x)$ is given by Eq. (9.21) in
Ref. \cite{jkg}.  The function $\xi(m_\chi)$ is given in Fig. 33
in Ref. \cite{jkg}; it quantifies the number of neutrino-induced
muons expected per annihilation event.  This depends in detail
on the annihilation products; e.g., $b \bar b$ or $c \bar c$
quarks, and/or $\tau^+ \tau^-$ lepton pairs.  Different WIMP
candidates in the mass ranges we are considering ($m_\chi
\lesssim 80$ GeV) will produce different branching fractions to
these annihilation products, and this will result in some
allowable range for $\xi(m_\chi)$.  For the WIMP masses we are
considering, $0.03 \lesssim \xi(m_\chi) \lesssim 0.13$; we use
$\xi=0.03$. The bound we compute here could be evaded
if the WIMP annihilated to $u\bar u$,
$d\bar d$, $s \bar s$, $e^+e^-$, and/or $\mu^+\mu^-$ pairs but
not $c \bar c$, $b \bar b$, nor $\tau^+\tau^-$ pairs.  However, 
we know of no models in which (nor any reason why) this would
occur. 
Eq.~(ref{eqn:indirectrate}) assumes that capture of WIMPs in the 
Sun from the Galactic halo is in equilibrium with their depletion 
by annihilation. 
As argued in Ref. \cite{jkg}, capture and annihilation will be in
equilibrium in just about any model in which the signal is
anywhere close to being detectable.  Since we are going to place 
upper limits to a WIMP-proton scattering cross section based on
current bounds, we may safely assume capture-annihilation
equilibration in our analysis.

The upper limit to the flux of neutrino-induced muons from 
the Sun (roughly $\lesssim10^{-2}~{\rm m}^{-2}~{\rm yr}^{-1}$)
{}from Super-Kamiokande~\cite{SK}, leads to the upper
limit to the cross section for WIMP-proton SD
scattering shown in Fig. \ref{fig:sdp} (limits from Baksan, 
MACRO and AMANDA lead to very similar results). The discrepancy 
of a couple of orders of magnitude between the region
disallowed by neutrino telescopes and the region implied by {\sc Dama} 
is much too large to be explained by the uncertainties and 
approximations inherent in our analysis, which might conceivably 
change our {\sc Dama} parameter space and/or our Super-Kamiokande 
limit by no more than factors of a few.

\begin{figure}[t]
\vskip -1.0cm
\centerline{\psfig{file=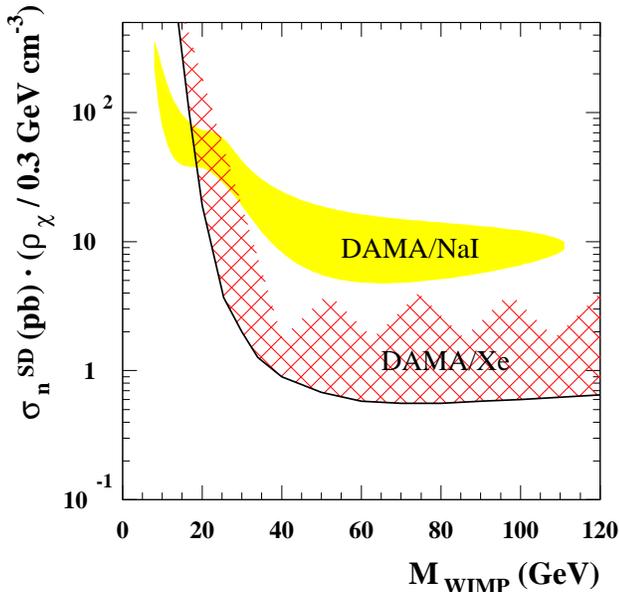,width=3.5in}}
\caption{The shaded region shows the parameter space (in WIMP
         mass versus SD WIMP-neutron cross section)
	 implied by the {\sc Dama} annual modulation under the
	 assumption that the signal is due exclusively to
	 SD WIMP-neutron scattering.  The solid curve
	 indicates the upper bound to the SD
	 WIMP-neutron cross section from the {\sc Dama} Xe detector.
	 Again the cross hatched region is excluded.}
\vskip -0.2cm
\label{fig:sdn}
\end{figure}

\section{A WIMP-NEUTRON INTERACTION?}

In the simplest odd-group model of the Na and I nuclei, the
spin with which a SD WIMP interacts is carried
exclusively by the unpaired proton.  In such a model, the WIMP
can undergo SD scattering with the nucleus only
through the WIMP-proton interaction, and not through any
WIMP-neutron interaction.  However, in more sophisticated
nuclear models, some small fraction of the spin is carried by
neutrons as well.  It is thus conceivable that the {\sc Dama}
modulation could have been caused by a SD WIMP-neutron interaction.
The point of this Section will be to show that this possibility 
is disfavored by the null results of WIMP searches with odd-neutron 
targets. 

In particular, the best current limit has been obtained 
{}from an experiment with enriched liquid Xe (99.5\% $^{129}$Xe)
performed as well by the {\sc Dama} collaboration
(we checked that a weaker limit is given by the {\sc Cdms} measurement 
with natural Ge, with only $\sim 8\%$ of the $^{73}$Ge isotope). 
We plot in Fig.~\ref{fig:sdn} the limit 
derived in Ref.~\cite{DAMAXe}
using again form factors from Ref.~\cite{NaIformfactor}, but 
assuming $a_{n}= - 0.85 \, a_{p}$ rather than 
$a_{p}= 0$, $a_{n}\neq 0$. The limit in the latter case should be 
even slightly lower than the one displayed. 
The limit from Xe excludes all WIMP masses larger than about 20
GeV, leaving just a tiny corner of the {\sc Dama} favored region.

\section{CONCLUSION}

The annual modulation observed in the {\sc Dama} low-energy bins has
been attributed to a WIMP with SI interactions
with nuclei, a possibility in contradiction with null
searches by {\sc Cdms}. However, this is {\it a priori} not the only
interpretation---the modulation could also be due most generally to 
a WIMP with SD interactions.  

We have shown here that the region of parameter space implied 
by the {\sc Dama} signal for a WIMP that undergoes SD scattering from 
protons is excluded in a model-independent way by null searches 
for a WIMP-induced energetic-neutrino flux from the Sun. 

The alternative hypothesis---that the modulation is due to WIMP 
scattering from neutrons---is excluded for $M_{\chi} \gsim 20$~GeV
by data taken with a Xe detector.  Although experimentally
viable, the $M_\chi\lesssim20$ GeV solution requires new physics 
beyond the minimal supersymmetric standard model \cite{opal}.  More
importantly, it requires that the WIMP-proton and WIMP-neutron
interactions differ by more than four orders of magnitude. Such
a large difference would be quite unusual.

The SD WIMP parameters compatible with the {\sc Dama} 
modulation derived here may be slightly changed by including particle 
and nuclear uncertainties or uncertainties in the local WIMP 
velocity distribution~\cite{veldist}, as well as by performing
a more accurate analysis on the raw data. However all these 
effects cannot account for the order-of-magnitude (or more)
discrepancies pointed out here. 

\medskip
We thank P. Belli and J. Edsj\"{o} for useful discussions.
This work was supported in part by NSF AST-0096023, NASA
NAG5-8506, and DoE DE-FG03-92-ER40701 and DE-FG03-88-ER40397.


\begin{thebibliography}{99}

\bibitem{jkg} G.~Jungman, M.~Kamionkowski, and K.~Griest,
     Phys. Rep. {\bf 267}, 195 (1996).

\bibitem{dirdet} 
     M.~W.~Goodman and E.~Witten, \physrevd{31}{1986}{3059};
     I. Wasserman, \physrevd{33}{1986}{2071}.

\bibitem{annmod} A.~Drukier, K.~Freese, and
     D.~Spergel, \physrevd{33}{1986}{3495};  K.~Freese, J.~Frieman,
     and A.~Gould, \physrevd{37}{1988}{3388}; K.~Griest,
     \physrevd{37}{1988}{2703}.


\bibitem{SOS} J.~Silk, K.~A. Olive, and M.~Srednicki,
     Phys.~Rev.~Lett. {\bf 55}, 257 (1985). 


\bibitem{DAMA} R.~Bernabei et al. ({\sc Dama} Collaboration),
     \plb{480}{2000}{23}.

\bibitem{CDMS}
     R.~Abusaidi et al. ({\sc Cdms} Collaboration), \prl{84}{2000}
     {5699}.

\bibitem{jkgs} M. Kamionkowski, K. Griest, G. Jungman, and
     B. Sadoulet, \physrevlett{75}{1995}{5174}.

\bibitem{katie} M. Kamionkowski and K. Freese,
     \physrevd{55}{1997}{1771}. 

\bibitem{DAMAXe} R.~Bernabei et al. ({\sc Dama} Collaboration),
     \plb{436}{1998}{379}.

\bibitem{bottino} A. Bottino, F. Donato, N. Fornengo, and
     S. Scopel, \physrevd{62}{2000}{056006}.

\bibitem{EngPitVog} J. Engel, S. Pittel, and P. Vogel,
     Int. J. Mod. Phys. E 1, 1 (1992).

\bibitem{BrhRos}
M.~Brhlik and L.~Roszkowski, \plb{464}{1999}{303}.

\bibitem{NaIformfactor} M. T. Ressell and D. J. Dean,
     \physrevc{56}{1997}{535}.


\bibitem{SK} A. Okada et al. (Super-Kamiokande collaboration),
     [\astroph{0007003}].

\bibitem{opal} G.~Abbiendi et al. (OPAL Collaboration), Eur.\
     Phys.\ J.\  {\bf C14}, 187 (2000).

\bibitem{veldist} M.~Kamionkowski and A.~Kinkhabwala,
     \physrevd{57}{1998}{3256}; P. Ullio and
     M. Kamionkowski, [hep-ph/0006183]; F. Donato, N. Fornengo,
     and S. Scopel, \astpar{9}{1998}{247};
     P. Belli et al., \physrevd{61}{2000}{023512}; 
     N. Evans, C. M. Carollo, and P. T. de Zeeuw, [\astroph{0008156}];
     A. M. Green, [\astroph{0008318}].

\end{thebibliography}
\end{document}